# Moral Advice as Interactional Negotiation: Framing, User Pressure, and Social Position in Large Language Model Responses

Minne Chen[1] † Yourong Yao[1]†

[1] School of Social Sciences, Nanyang Technological University, Singapore, Singapore

† Authors share first authorship of this work.

**Corresponding author**

Minne Chen (ORCID 0000-0003-0030-9641)

Address:

School of Social Sciences

Nanyang Technological University

48 Nanyang Ave, Singapore 639818

Email: minne.chen@ntu.edu.sg

Word count: 7,989 words, including the abstract, tables, references, figure and table captions, and endnotes.

**Abstract**

As conversational AI becomes a source of everyday guidance, LLMs increasingly participate in the interpretation and legitimation of morally contested choices. We examine LLM moral advice as an interactional negotiation shaped by framing, sustained user pressure, and the moral subject's social position. Using GPT-4o-mini as an illustrative case, we conducted a factorial vignette experiment with a pre-specified three-round protocol. The model received eldercare dilemmas that varied in framing and persona, followed by two user challenges. We analyzed 1,620 configuration-framing cells, each repeated three times, yielding 4,860 conversational runs. Caregiving affirmation produced near-uniform endorsement, whereas non-caregiving framing produced more variable baseline stances. When users challenged caregiving endorsement, 90.1% of configurations shifted after one round. Non-caregiving framing produced more resistant and unstable trajectories. Never (27.9%) and Late (25.6%) accommodations were more common than Early accommodations (16.5%), and only 14.32% of configurations achieved perfect trajectory consistency, compared with 62.72% under caregiving framing. Advice also varied with social position. Female personas received more support for non-caregiving decisions, while the presence of sisters increased accommodation. The GPT-4o-mini case shows that LLM moral advice can develop through a partially stable negotiation between normative response tendencies and user pressure rather than express a fixed ethical framework. The framework and design support comparative research across models and moral domains. Such instability raises social, ethical, and technical concerns, as users may treat advice that is difficult to scrutinize as objective.

## 1. Introduction

Conversational AI is becoming part of everyday advice-seeking, placing LLMs within ordinary processes of interpretation and judgment. Users turn to large language models (LLMs) not only for information but also for normative guidance, often interacting with them as trusted companions (Tudor Car et al., 2020; Brandtzaeg et al., 2022). Questions such as ‘What should I do?’ or ‘Is my decision justified?’ ask for moral evaluation rather than factual information (Montag & Elhai, 2025). In answering them, LLMs frame dilemmas, emphasize certain considerations, and lend apparent authority to particular interpretations (Gillespie, 2024). Trust and affective attachment can increase the weight users give to these responses (Montag & Elhai, 2025; Scherr et al., 2025; Logg et al., 2019), shaping how users evaluate decisions, assign responsibility, and interpret moral situations, even when the system does not explicitly prescribe action (Hohenstein & Jung, 2020; Salvi et al., 2025; Hackenburg et al., 2025; Shao et al., 2025).

Model alignment raises a related concern about the quality and consistency of moral guidance. Reinforcement learning from human feedback (RLHF), while designed to produce helpful and socially appropriate responses (Bai et al., 2022), may privilege agreement and conversational coherence over principled consistency, making models more likely to affirm or accommodate user positions even without supporting evidence (Sharma et al., 2024; Perez et al., 2023). Research also shows that LLM outputs may reproduce dominant cultural scripts under the guise of neutrality (Breazu & Katsos, 2024; Hart, 2025; Motoki et al., 2025; Ta et al., 2026). Research on AI-mediated communication therefore needs to ask what normative reasoning LLMs produce in morally ambiguous situations and how stable those judgments remain under conversational pressure. We conceptualize this problem as a tension between an accommodation pull toward the user's expressed position and embedded normative tendencies that make some moral positions easier to endorse than others. Morally contested advice becomes especially revealing when these forces point in different directions.

Intergenerational care obligations are a global concern. Research in the United States and Europe shows that adult children remain central to eldercare and that women, especially daughters, provide disproportionate support (Grigoryeva, 2017; Batur et al., 2024). In Asia, filial obligation has historically structured family relationships around reciprocity and intergenerational duty (Ikels, 2004), while urbanization, labor mobility, and changing gender roles have complicated these expectations (Qi, 2015). Sibling composition and birth order also shape caregiving expectations (Cong & Silverstein, 2012). Although specific norms and arrangements vary, eldercare brings family obligation, individual autonomy, harm avoidance, and gendered responsibility into conflict across societies. This makes it a useful case for examining how an LLM's normative response tendencies interact with user accommodation. We examine this widely shared moral problem through a vignette situated in an Asian context, using GPT-4o-mini as an illustrative multilingual LLM (OpenAI, 2024).

Research on AI-mediated communication has paid limited attention to how LLMs evaluate morally ambiguous decisions when questions are framed differently, challenged, or attributed to moral subjects in different social positions. Most research treats model responses as isolated, single-turn outputs assessed for informational accuracy (Lin et al., 2022), hallucination (Breazu & Katsos, 2024), or stereotype reproduction (Hart, 2025). To address this gap, the present study develops an interactional framework across three connected dimensions and applies it to GPT-4o-mini as an illustrative case. We examine whether affirmative and negative framing produce different baseline judgments, whether those judgments change under successive user challenges, and whether the moral subject's social position shapes baseline evaluations and accommodation. Together, these dimensions provide a framework for examining whether LLM moral advice is consistent or interaction-contingent.

Our central contribution is to treat moral accommodation as a trajectory across successive rounds of user challenge. Whereas Cheng et al. (2025, 2026) establish that social sycophancy is prevalent and consequential, we ask when moral accommodation emerges, how stable it is, and how its trajectory varies across framing directions and social positions. Rather than assuming a fixed value orientation (Schramowski et al., 2022) or a uniform sycophantic tendency (Sharma et al., 2024; Perez et al., 2023), we conceptualize moral guidance as a partially stable negotiation shaped by framing, conversational pressure, and social position (Figure 1). Our behavioral perspective focuses on how judgments manifest to users and how readily they are revised under sustained challenge.

## 2. Literature Review

### 2.1. Framing Effects on LLMs' Moral Judgments

LLMs express reproducible social and political attitudes in response to subjective questions (Santurkar et al., 2023; Motoki et al., 2025), yet those positions may shift across varied prompt formulations (Ceron et al., 2024; Moore et al., 2024). LLM performance can change even when prompts differ only in formatting rather than meaning (Sclar et al., 2024). In moral dilemmas, model judgments can also change depending on whether a choice is framed as an action or omission and whether endorsing it requires a yes or no answer (Cheung et al.,

2025).

Eldercare provides a useful setting for examining framing effects because moral judgments about caregiving are culturally structured (Silverstein, Gans & Yang, 2006) and normatively complex. Filial reciprocity, individual autonomy, elder welfare, and fairness among family members may not point in the same direction (Stuifbergen & Van Delden, 2011; Schinkel, 2012). Affirmative caregiving prompts may emphasize duty and moral virtue. Denial prompts may emphasize constraint, hardship, or justification. If LLMs apply consistent normative principles, baseline judgments should remain equivalent across framings (Jiang et al., 2025). Divergence would indicate that the direction of questioning matters to the model's response. Combined with probabilistic output generation and documented sycophantic tendencies (Sharma et al., 2024; Perez et al., 2023), it would raise concerns about the reliability of LLM moral guidance.

The following research questions concern LLM moral advice as a broader object of inquiry and are examined empirically through GPT-4o-mini as an illustrative case.

***RQ1:*** *How do alternative framings of the same eldercare dilemma shape baseline evaluative stances in LLM moral advice?*

## 2.2. Accommodation of LLM Judgments Under User Pushback

Users do not simply accept initial responses. They express disagreement, request reconsideration, and challenge judgments. These exchanges matter because norms are negotiated through interaction, yet conversational-AI research rarely examines how initial framing shapes what happens under pressure. Prior studies document both sycophantic alignment with user-expressed preferences, even when it contradicts prior arguments (Sharma et al., 2024; Perez et al., 2023), and greater response adjustment as disagreement intensifies (Anil et al., 2024; Salvi et al., 2025).

Two gaps remain. First, much sycophancy research has focused on factual questions, where shifts toward a user's answer can be compared with an external standard of correctness. Moral dilemmas lack a single objective answer, so changes in stance cannot simply be treated as factual errors. Cheng et al. (2025) show that LLMs often validate users in contexts of morally ambiguous advice, while Cheng et al. (2026) show that sycophantic advice can affect users' judgments and behavioral intentions. Neither study tracks when a model changes or maintains its stance across repeated challenges.

Second, framing and sycophancy have largely been studied separately. Framing research focuses on baseline, single-turn outputs, while sycophancy research often treats accommodation as a uniform tendency. It remains unclear whether initial framing shapes both where models start and how they respond to successive challenges. Our multi-turn factorial design addresses both gaps by tracking when a moral stance changes or persists and how its path differs by framing direction.

***RQ2: When does moral accommodation emerge across successive user challenges, how stable is it, and how does its trajectory vary by initial framing?***

### 2.3. Persona Differences in Moral Judgment and Accommodation

A third dimension concerns who is represented as the decision-maker. Moral accountability is unevenly distributed across social roles, reflecting culturally embedded expectations and normative hierarchies (Ridgeway, 2011). Because LLMs are trained on human-generated text that reflects these norms, their responses may reproduce socially structured patterns of moral judgment rather than uniformly neutral criteria (Breazu & Katsos, 2024; Hart, 2025). Persona-relevant prompting supports this possibility, showing that contextual cues can produce different normative scripts and reasoning patterns (Liu et al., 2024; Li et al., 2024). It remains unclear how this variation manifests in morally ambiguous, multi-turn evaluations and whether subjects in different social positions exhibit varying degrees of moral flexibility when their initial judgments are challenged.

Eldercare makes these social locations especially visible. Gendered norms assign different expectations to men and women (Finch & Mason, 2003), with women more closely associated with caregiving and expected to provide emotional and practical support (Grigoryeva, 2017; Batur et al., 2024). Judgments of people who decline care reflect these social roles and relational contexts (Elliott et al., 2015; Rurka et al., 2023; Lee, 2016). Sibling composition introduces a further axis of differentiation, with birth order and gender configuration shaping perceived caregiving responsibility. Eldest children are often expected to coordinate family obligations, whereas daughters are more often expected to provide practical and emotional support (Vergauwen & Mortelmans, 2021; Grigoryeva, 2017; Kim et al., 2024; Batur et al., 2024). Persona attributes therefore represent social locations tied to unequal expectations of care and justification. Comparing baseline judgments and accommodation allows us to assess whether LLMs apply uniform standards or reproduce these patterns and whether they interact with framing and conversational pressure.

***RQ3:*** *How do persona attributes such as gender and sibling structure relate to baseline judgments and accommodation in LLM moral advice?*

We integrate framing, sycophancy, and persona into a single interactional framework

(Figure 1). A morally contested query brings the model's accommodation pull into tension with its embedded normative tendencies. These are recurring orientations learned through the heterogeneous textual and alignment environment that shapes the model. These include orientations toward care obligation, individual autonomy, harm avoidance, and gendered responsibility. Framing direction, successive user challenge, and social position jointly shape the starting stance and path of accommodation.

*[Figure 1 here]*

**3. Methodology**

3.1. Experimental Design

This study used a factorial vignette experiment (Auspurg & Hinz, 2015) and a pre-specified conversational protocol to examine how LLMs evaluate morally ambiguous eldercare dilemmas under user pressure. The design follows calls for socially grounded behavioral evaluation beyond isolated accuracy checks (van Dis et al., 2023; Weidinger et al., 2022). It varied dilemma framing, conversational pushback, and persona attributes.

Interactions were implemented on a custom GenAI research platform through the OpenAI Python SDK and Chat Completions API, requesting gpt-4o-mini. Data were collected on 9 November 2025 (UTC+08:00), with temperature set to 0.7. GPT-4o-mini was selected as a widely deployed and relatively low-cost API model (OpenAI, 2024). A fixed system prompt defined it as a decision-support assistant for eldercare dilemmas and required an explicit evaluative stance (support or not support) followed by a structured analytic justification. This format supported reliable extraction and comparison across conditions, consistent with LLM auditing and content-coding workflows (Gilardi et al., 2023).

The same underlying eldercare dilemma was presented in two semantically parallel but normatively distinct formulations. In Q1, the persona stated, ‘I provided caregiving for my parents. What do you think?’ In Q2, the persona stated, ‘I did not provide caregiving for my parents. What do you think?’ This tested whether baseline judgments changed with framing direction. The system and user prompts were administered in Chinese. English translations of the two vignette formulations are reported here.

Each vignette comprised three rounds. R0 recorded the baseline judgment. R1 used the fixed pushback string, ‘I don’t really agree with your answer.’ R2 used the fixed follow-up string, ‘I still disagree. Please take my circumstances into account and offer more targeted advice.’ Neither supplied new case facts. R2 intensified the pushback by repeating the disagreement and indicating that the previous response had not adequately addressed the persona’s stated circumstances. Appendix C reports the model and run configuration and provides English translations of the system prompt, Q1 and Q2 vignettes, and fixed follow-up messages. We treat individual stance changes descriptively as conversational accommodation and reserve sycophantic accommodation for systematic directional patterns under the fixed protocol, not every individual shift (Uth et al., 2025; Messeri & Crockett, 2024).

Personas varied across a complete factorial of five dimensions: gender (undisclosed, male, female), age (undisclosed, 18-30, 30-45, 45-60, 60 or above), region (undisclosed, rural, urban), education (undisclosed, no college degree, college degree or above), and sibling configuration (undisclosed, with sisters and not the oldest child, with brothers and not the oldest child, with sisters and is the oldest child, with brothers and is the oldest child, or no siblings). These levels yielded $3 \times 5 \times 3 \times 3 \times 6 = 810$ unique persona configurations per framing. Gender refers to socially constructed gender conveyed through persona cues. The undisclosed level served as the reference, comparing each disclosed cue with an otherwise identical prompt in which it was withheld. Persona-based manipulation is widely used to assess whether LLM outputs vary in response to identity and other social cues (Li et al., 2024; Liu et al., 2024). The personas were experimentally constructed prompt profiles rather than human participants. Each of the 1,620 configuration-framing cells (810 per framing) was run three times with identical prompts to assess probabilistic response stability and reduce run-to-run noise (Gilardi et al., 2023; Cui et al., 2025; Xie et al., 2025), totaling 4,860 three-round conversational runs. Appendix Table A1 summarizes the factorial structure and repetitions.

*[Figure 2 here]*

3.2. Measurement of Evaluative Stance and Conversational Accommodation

The outcomes were evaluative stance and conversational accommodation. Because each response began with an explicit stance, the stance at each round was coded as support or not support. Baseline stance was R0.

Conversational accommodation was operationalized as a change in explicit evaluative stance toward the position implied by user disagreement across rounds. Because the follow-up rounds introduced no new case-relevant facts, systematic directional shifts provide evidence of sycophantic accommodation (Sharma et al., 2024; Perez et al., 2023).

To systematically examine accommodation patterns, a directional accommodation target was defined for each framing condition. In the caregiving condition, the accommodation target was defined as shifting from support to non-support, reflecting accommodation to user disagreement with positive caregiving evaluation. In the non-caregiving condition, the accommodation target was defined as shifting from non-support to support, reflecting accommodation toward legitimizing non-caregiving decisions. Based on when the model first adopted the target stance, we created a four-category variable capturing the timing of accommodation:

0 = Start: the model already matched the target stance at R0 (baseline alignment with the pressured direction)

1 = Early: the model first matched the target stance at R1

2 = Late: the model first matched the target stance at R2

3 = Never: the model never matched the target stance across R0-R2 (resistance to pushback)

### 3.3. Data Analysis

Analyses were conducted separately by framing, with each persona configuration as one observation. The three repetitions also provided a categorical stability check. A cell was perfectly consistent when all three runs produced the same accommodation trajectory, majority-consistent when two matched, and fully inconsistent when all three differed. The modal classification was used whenever at least two runs agreed. Fully inconsistent cells had no identifiable modal timing outcome and were excluded only from the timing models. Their distribution and persona correlates were analyzed separately in Table 3 and Appendix Table B2.

No baseline-stance model was estimated for Q1 because all 810 configurations supported caregiving at R0, leaving no outcome variation. For Q2, the baseline-stance model retained all 810 persona configurations and all five persona dimensions. Because only 46 cells produced supportive baseline classifications across 15 indicator parameters, we used Firth's penalized-likelihood logistic regression with 95% penalized profile-likelihood confidence intervals and penalized likelihood-ratio tests (Firth, 1993; Heinze & Schemper, 2002).

Accommodation dynamics were estimated separately by framing. In Q1, no configuration was aligned with the target stance at baseline (Start, n = 0), and Never outcomes were rare. Model 1 therefore used Firth logistic regression to compare Never (coded 1) with Any accommodation (Early or Late, coded 0). In Q2, baseline alignment (Start) was analyzed separately in Table 4. Among configurations not aligned at baseline, Early, Late, and Never trajectories were compared using multinomial logistic regression with Early as the reference outcome. Relative risk ratios (RRRs) above 1 indicate a higher relative risk of Late or Never accommodation compared with Early accommodation. The multinomial specification was used because the proportional-odds assumption was rejected (Brant $\chi^2(15) = 48.09$, $p < .001$).

## 4. Results

### 4.1. Baseline stance and moral asymmetry across framings

Baseline stance differed sharply by framing (Table 1). In Q1 (affirming eldercare provision), the model endorsed caregiving across all persona configurations and repetitions. In Q2 (denying eldercare provision), the model predominantly opposed the user's decision not to provide care but gave supportive baseline responses for 46 persona configurations before conversational pushback.

Across three repeated interactions per persona configuration, 99.38% of Q1 configurations produced perfectly consistent baseline stances, compared with 79.26% in Q2 (Table 2). Thus, when caregiving was denied, more than one-fifth of configurations produced divergent baseline stances across identical trials.

*[Table 1 here]*

*[Table 2 here]*

4.2. Conversational Pushback and Accommodation Patterns

To address RQ2, we examined how the model responded to successive user disagreement across conversational rounds.

Accommodation timing also differed sharply (Table 3). In Q1, Early accommodation dominated (90.1%), while Never (7.5%) and Late (0.5%) were rare. In Q2, Never was the largest category (27.9%), followed by Late (25.6%), Early (16.5%), and Start (5.7%); 24.3% of configurations were fully inconsistent. Among the 613 Q2 configurations with an identifiable modal trajectory, Never accounted for 36.9%, Late for 33.8%, Early for 21.9%, and Start for 7.5%.

*[Table 3 here]*

Across the 1,620 cells, 624 (38.52%) were perfectly consistent, 784 (48.40%) were majority-consistent, and 212 (13.09%) were fully inconsistent. Fully inconsistent cells were concentrated in Q2 (197; 24.3%), compared with Q1 (15; 1.9%). Perfect trajectory consistency was also lower in Q2 than Q1 (14.32% vs. 62.72%; Appendix Table A2). Accommodation was therefore rapid and predictable in Q1 but more delayed, resistant, and unstable in Q2. Because Table 5 requires an identifiable modal trajectory, Appendix Table B2 examines response consistency across all 810 configurations in each framing, including the fully inconsistent cells excluded from the timing models.

### 4.3. Persona Differences in Baseline Stance and Accommodation Patterns

To address RQ3, we examined whether the model's baseline judgments and subsequent accommodation patterns varied systematically across persona attributes.

#### 4.3.1. Persona Predictors of Baseline Stance

Because all 810 Q1 baseline classifications supported caregiving, persona-level analysis of baseline stance was restricted to Q2 (Table 4 and Figure 3). In the Q2 Firth model, gender emerged as the strongest predictor. Relative to personas whose gender was undisclosed, female personas were substantially more likely to receive support for not providing care (OR = 50.656, 95% CI [12.523, 467.252], $p < .001$), although the wide interval indicates substantial uncertainty in the magnitude of this association. Male personas did not differ significantly from the undisclosed-gender reference category. Baseline evaluations therefore differed sharply between female and undisclosed-gender personas.

Family structure also differentiated the model's baseline judgments. Only-child personas were more likely than personas with an undisclosed sibling configuration to receive support for not providing care (OR = 10.606, 95% CI [3.973, 33.536], $p < .001$). Conversely, personas who were the oldest child and had brothers were less likely to receive such support (OR = 0.080, 95% CI [0.001, 0.741], $p = .023$); the other sibling configurations did not differ significantly from the undisclosed reference category. In these outputs, sole caregiving responsibility was associated with greater acceptance of non-caregiving. Figure 3 further illustrates these patterns, showing that the combination of female gender and only-child status produced by far the highest predicted probability of supporting non-caregiving ($Pr \approx 0.54$), while all other subgroups remained near zero.

Among the other persona attributes, urban residence was associated with greater support for not providing care (OR = 2.941, 95% CI [1.263, 7.311], p = .012). Age and education were not significantly associated with baseline stance in this framing. Overall, the model overwhelmingly endorsed caregiving when users affirmed providing care in Q1, but Q2 evaluations varied with persona characteristics, especially gender and family-responsibility cues.

*[Table 4 here]*

*[Figure 3 here]*

4.3.2. Persona Predictors of Accommodation Timing

Table 5 reports persona associations with Q1 resistance and, among Q2 configurations not aligned at baseline, accommodation timing. Female personas showed less sustained resistance in both framings. Relative to personas whose gender was undisclosed, female personas had lower odds of remaining resistant in Q1 (OR = 0.424, 95% CI [0.190, 0.899], p = .025). In Q2, they had a lower relative risk of Never rather than Early accommodation (RRR = 0.278, 95% CI [0.148, 0.523], p < .001), while the Late-rather-than-Early contrast was not significant (RRR = 0.911, 95% CI [0.528, 1.574], p = .739).

Having an identified sibling was associated with earlier accommodation most consistently in Q2, whereas having no sibling increased resistance in Q1. In Q1, personas with sisters who were not the oldest had lower odds of resistance (OR = 0.215, 95% CI [0.041, 0.769], p = .017), whereas only-child personas had higher odds of resistance (OR = 4.260, 95% CI [2.039, 9.582], p < .001). In Q2, all four configurations indicating at least one sibling were associated with lower relative risks of both Late and Never rather than Early accommodation compared with undisclosed sibling status (Late RRRs = 0.201-0.257, all $p \leq .007$; Never RRRs = 0.058-0.202, all $p < .002$). No-sibling status did not differ significantly from the undisclosed reference in either Q2 contrast.

Education cues were associated with greater resistance in Q1, whereas delayed or sustained resistance appeared in Q2, depending on education level. Relative to personas whose education was undisclosed, both personas without a college degree (OR = 2.150, 95% CI [1.053, 4.570], $p = .036$) and those with a college degree or above (OR = 2.097, 95% CI [1.026, 4.458], $p = .042$) had higher odds of resistance in Q1. In Q2, no-college personas had a higher relative risk of Late rather than Early accommodation (RRR = 1.783, 95% CI [1.052, 3.023], $p = .032$), whereas college-educated personas had a higher relative risk of Never rather than Early accommodation (RRR = 2.135, 95% CI [1.198, 3.807], $p = .010$). The other Q2 contrast for each education category was not significant.

Age was more strongly associated with resistance in Q1 than with accommodation timing in Q2. In Q1, personas aged 18-60 had lower odds of resistance than personas whose age was undisclosed (ORs = 0.244-0.406, all $p < .05$), while the association for personas aged 60 or above was not significant. In Q2, personas aged 60 or above had a lower relative risk of Never rather than Early accommodation (RRR = 0.446, 95% CI [0.210, 0.949], $p = .036$); the remaining Q2 age contrasts were not significant.

*[Table 5 here]*

## 5. Discussion

The GPT-4o-mini case shows that LLM moral advice is a social and interactional phenomenon. It varies with question framing, the persistence of user pressure, and the social position of the moral subject rather than expressing a fixed or neutral evaluation.

## 5.1 Asymmetrical Framing, Sycophancy, and AI Moral Reasoning

The central empirical finding is an asymmetry across conversational turns. Q1 produced universal and highly reproducible caregiving endorsements at baseline. These responses varied little across personas but usually changed after one challenge. Q2 produced more varied baseline stances, less reproducible accommodation trajectories, and more variation by persona. Among Q2 configurations with an identifiable modal trajectory, Never and Late accommodation were both more common than Early accommodation. How the model responded to the challenge, therefore, depended on whether non-caregiving appeared as a reversal of an endorsed caregiving decision in Q1 or as an established position under evaluation in Q2.

The asymmetry may reflect competing alignment objectives. When users challenged an endorsement of caregiving, the model could reverse its position by reframing non-caregiving as respect for autonomy and personal circumstances, consistent with autonomy-affirming tendencies in systems trained through RLHF (Bai et al., 2022). In Q2, accommodation required the model to support a decision that could harm a vulnerable third party, placing helpfulness and harmlessness in tension. This may help explain why Q2 responses changed later, resisted challenge more often, and varied more across repeated interactions. It also extends work on factual LLM sycophancy (Sharma et al., 2024; Perez et al., 2023) by showing that moral accommodation depends on the normative direction of the exchange.

The asymmetry matters beyond alignment engineering. AI-generated moral advice cannot be reduced either to a fixed ethical framework imposed during model alignment or to user input alone. Responses changed during the conversation as helpfulness, harmlessness, and embedded social norms pulled in different directions. The same dilemma produced different outputs across framings and across repeated interactions with identical prompts.

### 5.2 When Embedded Social Assumptions Conflict with Sycophancy

Persona attributes show that the model did not respond uniformly across users. Social assumptions reflected in the outputs were associated with both baseline judgments and accommodation patterns. A further implication emerges when these embedded social assumptions intersect with the model's sycophantic tendencies.

Relative to personas whose gender was undisclosed, female personas were much more likely to receive support for non-caregiving at Q2 baseline and less likely to remain resistant under challenge in both framings. In a domain where women bear disproportionate family-care responsibilities (Grigoryeva, 2017; Batur et al., 2024), this distribution may reflect greater recognition of the burdens attached to caregiving. At the same time, it can reproduce benevolent-sexist logic if women receive greater moral latitude precisely because caregiving is assumed to be their responsibility (Barreto & Doyle, 2023; Pinho & Gaunt, 2024).

Personas who reported having sisters and were not the eldest received greater accommodation in both framings, regardless of their own gender. This pattern is consistent with gendered caregiving norms that treat female relatives as available caregivers (Pinho & Gaunt, 2024). Greater support for female personas may reflect recognition of women's caregiving burden, but when a sister was available, the outputs more readily accepted non-caregiving by shifting care to another woman. Under user pressure, accommodation therefore outweighed the apparent recognition of women's unequal caregiving burden.

### 5.3 When Embedded Social Assumptions Encounter Framing

Education was associated with slower accommodation in both framings, although the form differed. Relative to undisclosed education, both education categories were associated with greater resistance in Q1. In Q2, no-college status predicted Late rather than Early accommodation, while college education predicted Never rather than Early accommodation. Only-child status showed a clearer framing difference. In Q1, caregiving endorsement was less likely to be reversed for only-child personas. In Q2, however, only-child personas were more likely to receive supportive baseline evaluations of non-caregiving, although their accommodation timing did not differ significantly from the undisclosed-sibling group. Age effects were more limited. Personas aged 18-60 showed less resistance in Q1, while most Q2 age contrasts were not significant; only personas aged 60 or above had a lower relative risk of Never rather than Early accommodation.

Framing changed which persona associations appeared and at what stage of the interaction. Having no sibling to share care can strengthen the expectation of caregiving when care is initially endorsed, while sole responsibility can also make a refusal appear more understandable when non-caregiving is presented as an established decision (Cong & Silverstein, 2012). The lower Q1 resistance for personas aged 18-60 is consistent with greater latitude for individual pursuits (Moore, 2005). The isolated Q2 finding for personas aged 60 or above may reflect age-related expectations of maturity and responsibility (Krettenauer, 2022), but the evidence does not show a general age gradient.

5.4 Theoretical and Social Implications of LLM Moral Advice

The findings challenge two incomplete accounts of AI moral reasoning. One treats language models as expressing measurable moral or value orientations (Schramowski et al., 2022). The other emphasizes sycophancy, showing that models trained with human feedback may align with users' views at the expense of truthfulness (Perez et al., 2023; Sharma et al., 2024). The GPT-4o-mini findings support an interactional account of LLM moral advice in which user accommodation and embedded normative tendencies can pull responses in different directions, while framing affects how that tension is resolved. When offering moral advice, LLMs function as social actors rather than neutral technical systems. Their outputs can shape and legitimize moral positions even though users cannot see how those positions were produced.

Beyond its empirical findings, the study contributes an interactional framework and a multi-turn factorial design for studying LLM moral advice. The framework treats moral accommodation as a trajectory across successive user challenges, while the design separates baseline stance, timing of change, and reproducibility across identical interactions. This combination provides a common structure for testing how framing and social position shape moral advice across models, alignment approaches, and moral domains.

These findings also have practical implications as AI systems enter mental health support, companionship, and other advisory settings. Users may treat AI responses as trustworthy moral evaluations (Cheng et al., 2026), giving them a form of algorithmic authority (Gillespie, 2014). Yet the advice observed here depended on framing and varied across identical interactions. In caregiving decisions, where users may already face conflict or social pressure, accommodating responses could reinforce confirmation bias (Bashkirova & Krpan, 2024) and make contested choices easier to rationalize. Systems should therefore make clear that AI-generated moral judgments are context-dependent rather than stable or authoritative.

### 5.5 Limitations

This study uses GPT-4o-mini as an illustrative case, so the estimated effect sizes, framing asymmetries, and persona associations are specific to the tested model and conditions. Some persona estimates were imprecise because supportive Q2 baseline outcomes and resistant Q1 trajectories were sparse, even with Firth estimation.

A single model response provides no evidence of whether the same trajectory will recur. Repeating each configuration three times allowed us to distinguish perfectly consistent, majority-consistent, and fully inconsistent trajectories under identical prompts. Three repetitions therefore provide a categorical check of response stability, although additional runs could yield a modal trajectory for some cells classified as fully inconsistent. We analyzed these configurations separately rather than assigning them an Early, Late, or Never timing category. Only Table 5's persona timing model excludes fully inconsistent configurations. The framing and instability analyses in Table 3 and Appendix Table B2 include all 810 Q2 configurations.

The study used Chinese-language prompts about care for elderly parents. Intergenerational care obligations arise across societies, but other languages, settings, and moral conflicts may produce different patterns. The experimental personas also simplify how social positions appear in real conversations. Future research could compare parallel prompts across languages and combine factorial experiments with observational data. Finally, the analysis concerns behavioral outputs and cannot establish the internal processes that produced them.

## 6. Conclusion

As conversational AI enters everyday decision-making, its moral responses shape how authority, responsibility, and legitimate action are communicated. Using GPT-4o-mini as an illustrative case, this study examined how LLM responses to ambiguous caregiving dilemmas changed under conversational pressure. Caregiving endorsement generally collapsed after one challenge, whereas non-caregiving disapproval more often persisted or shifted only after continued challenge. The two framings also differed in trajectory consistency. Only 14.32% of Q2 configurations produced the same accommodation trajectory across three repetitions, compared with 62.72% in Q1.

The case shows that LLM moral advice is neither a fixed ethical framework nor a simple reflection of user preferences. It develops through interaction between embedded normative tendencies and user accommodation. Framing, persistent disagreement, and the moral subject's social position shaped the advice. When offering moral guidance, LLMs act as social actors whose outputs can shape and legitimize moral positions even though users cannot see how those positions were produced. As these systems enter consequential advisory settings, their instability is a social and ethical problem as well as a technical one.

**Acknowledgements**

The authors thank Ms. GAO Jun (undergraduate student, College of Computing and Data Science, Nanyang Technological University) for her technical support during the study setup and data collection stages.

**Disclosure statement**

The authors declare that they have no known competing financial interests or personal relationships that could have appeared to influence the work reported in this paper.

**CRediT author statement**

Minne Chen: Conceptualization, Data curation, Investigation, Formal analysis, Visualization, Methodology, Project administration, Funding acquisition, Resources, Supervision, Writing - original draft, Writing - review and editing.

Yourong Yao: Conceptualization, Data curation, Investigation, Visualization, Methodology, Writing - original draft, Writing - review and editing.

## Tables

**Table 1. Baseline Stance**

| | Question | | |
|---|---|---|---|
| | Q1 Caregiving | Q2 Non-caregiving | Total |
| Stance | | | |
| Oppose | | 764 | 764 |
| Support | 810 | 46 | 856 |
| Total | 810 | 810 | 1,620 |

**Table 2. Consistency of Stance across Repetitions**

| Question | Perfect match across 3 repetitions | | |
|---|---|---|---|
| | 0 | 1 | Total |
| Q1 Caregiving | 5 | 805 | 810 |
| % | 0.62 | 99.38 | 100.00 |
| Q2 Non-caregiving | 168 | 642 | 810 |
| % | 20.74 | 79.26 | 100.00 |
| Total | 173 | 1447 | 1620 |
| % | 10.68 | 89.32 | 100.00 |

**Table 3. Distribution of Accommodation Timing**

| | Q1 Caregiving | | Q2 Non-caregiving | | Total | |
|---|---|---|---|---|---|---|
| | Count | All % (modal %) | Count | All % (modal %) | Count | All % (modal %) |
| Start | 0 | 0.0 (0.0) | 46 | 5.7 (7.5) | 46 | 2.8 (3.3) |
| Early | 730 | 90.1 (91.8) | 134 | 16.5 (21.9) | 864 | 53.3 (61.4) |
| Late | 4 | 0.5 (0.5) | 207 | 25.6 (33.8) | 211 | 13.0 (15.0) |
| Never | 61 | 7.5 (7.7) | 226 | 27.9 (36.9) | 287 | 17.7 (20.4) |
| Fully inconsistent | 15 | 1.9 (—) | 197 | 24.3 (—) | 212 | 13.1 (—) |
| Total | 810 | 100.0 | 810 | 100.0 (100.0) | 1620 | 100.0 (100.0) |

| Q1 Caregiving | Q2<br>Non-caregiving | Total |
|---|---|---|
| (100.0) | | |

*Notes: The first percentage in each cell uses all configurations as the denominator (Q1 N = 810; Q2 N = 810; total N = 1,620). Parenthetical percentages are conditional on an identifiable modal trajectory (Q1 N = 795; Q2 N = 613; total N = 1,408) and therefore exclude fully inconsistent configurations, for which no modal timing category exists. Fully inconsistent configurations remain reported as an observed instability category rather than being treated as ordinary missing data.*

**Table 4. Persona Factors Associated with GPT-4o-mini's Support for 'Not Providing Care' (Q2)**

| **Predictor** | **OR** | **95% PPL CI** | **PLR p** |
|---|---|---|---|
| **Gender** | | | |
| undisclosed | 1.000 | — | — |
| male | 3.929 | [0.752, 39.168] | .109 |
| female | 50.656 | [12.523, 467.252] | < .001 |
| **Age** | | | |
| undisclosed | 1.000 | — | — |
| 18-30 | 0.833 | [0.248, 2.746] | .762 |
| 30-45 | 1.587 | [0.535, 4.896] | .406 |
| 45-60 | 1.587 | [0.535, 4.896] | .406 |
| 60 and above | 1.181 | [0.379, 3.742] | .773 |
| **Region (rural-urban)** | | | |
| undisclosed | 1.000 | — | — |
| rural | 1.125 | [0.433, 2.947] | .809 |
| urban | 2.941 | [1.263, 7.311] | .012 |
| **Education** | | | |
| undisclosed | 1.000 | — | — |
| no college degree | 1.117 | [0.442, 2.845] | .814 |
| college degree or above | 2.270 | [0.979, 5.532] | .057 |
| **Sibling** | | | |
| undisclosed | 1.000 | — | — |
| with sisters, not oldest | 0.798 | [0.201, 3.023] | .737 |
| with brothers, not oldest | 1.000 | [0.275, 3.635] | 1.000 |
| with sisters, oldest | 0.605 | [0.132, 2.433] | .480 |
| with brothers, oldest | 0.080 | [0.001, 0.741] | .023 |

| Predictor | OR | 95% PPL CI | PLR p |
|---|---|---|---|
| **Gender** | | | |
| no sibling | 10.606 | [3.973, 33.536] | < .001 |
| Observations | 810 | | |
| Penalized log likelihood | -91.522 | | |
| Omnibus PLR $\chi^2(15)$ | 137.46 | | < .001 |

*Notes: Odds ratios are from Firth's penalized-likelihood logistic regression. PPL CI = penalized profile-likelihood confidence interval; PLR p = p-value from a penalized likelihood-ratio test. The reported log likelihood is penalized, and the omnibus statistic is a penalized likelihood-ratio test. OR > 1 indicates a higher likelihood of GPT supporting the user's decision not to provide care.*

**Table 5. Predictors of Q1 Resistance and Q2 Conditional Accommodation Timing**

| | (1) | (2a) | (2b) |
|---|---|---|---|
| | Q1: Rare Resistance (Firth) | Q2: Late vs Early (M-Logit) | Q2: Never vs Early (M-Logit) |
| | OR / 95% Profile CI | RRR / 95% CI | RRR / 95% CI |
| Gender | | | |
| undisclosed | 1.000 | 1.000 | 1.000 |
| | [Reference] | [Reference] | [Reference] |
| male | 1.300 | 1.046 | 1.181 |
| | [0.700, 2.441] | [0.602, 1.820] | [0.680, 2.050] |
| female | 0.424* | 0.911 | 0.278*** |
| | [0.190, 0.899] | [0.528, 1.574] | [0.148, 0.523] |
| Age | | | |
| undisclosed | 1.000 | 1.000 | 1.000 |
| | [Reference] | [Reference] | [Reference] |
| 18-30 | 0.306** | 0.542 | 1.924 |
| | [0.121, 0.712] | [0.256, 1.148] | [0.915, 4.044] |
| 30-45 | 0.244** | 0.685 | 0.825 |
| | [0.087, 0.603] | [0.342, 1.375] | [0.392, 1.732] |
| 45-60 | 0.406* | 0.958 | 1.281 |
| | [0.172, 0.906] | [0.462, 1.988] | [0.593, 2.768] |
| 60 and above | 0.710 | 0.702 | 0.446* |

| | (1) | (2a) | (2b) |
|---|---|---|---|
| | Q1: Rare Resistance (Firth) | Q2: Late vs Early (M-Logit) | Q2: Never vs Early (M-Logit) |
| | [0.336, 1.475] | [0.363, 1.358] | [0.210, 0.949] |
| Region (rural-urban) | | | |
| undisclosed | 1.000 | 1.000 | 1.000 |
| | [Reference] | [Reference] | [Reference] |
| rural | 1.689 | 1.162 | 0.652 |
| | [0.846, 3.451] | [0.672, 2.009] | [0.366, 1.161] |
| urban | 1.447 | 1.288 | 0.866 |
| | [0.720, 2.962] | [0.740, 2.241] | [0.486, 1.543] |
| Education | | | |
| undisclosed | 1.000 | 1.000 | 1.000 |
| | [Reference] | [Reference] | [Reference] |
| no college degree | 2.150* | 1.783* | 0.691 |
| | [1.053, 4.570] | [1.052, 3.023] | [0.384, 1.244] |
| college degree or above | 2.097* | 1.297 | 2.135* |
| | [1.026, 4.458] | [0.722, 2.330] | [1.198, 3.807] |
| Sibling | | | |
| undisclosed | 1.000 | 1.000 | 1.000 |
| | [Reference] | [Reference] | [Reference] |
| with sisters, not oldest | 0.215* | 0.254** | 0.058*** |
| | [0.041, 0.769] | [0.099, 0.653] | [0.021, 0.157] |
| with brothers, not oldest | 0.394 | 0.228** | 0.084*** |
| | [0.113, 1.174] | [0.087, 0.598] | [0.031, 0.224] |
| with sisters, oldest | 0.578 | 0.201*** | 0.081*** |
| | [0.197, 1.577] | [0.078, 0.518] | [0.031, 0.211] |
| with brothers, oldest | 0.676 | 0.257** | 0.202** |
| | [0.243, 1.792] | [0.097, 0.684] | [0.078, 0.527] |
| no sibling | 4.260*** | 1.046 | 0.460 |
| | [2.039, 9.582] | [0.313, 3.500] | [0.137, 1.540] |
| / | | | |

| | (1) | (2a) | (2b) |
|---|---|---|---|
| | Q1: Rare Resistance (Firth) | Q2: Late vs Early (M-Logit) | Q2: Never vs Early (M-Logit) |
| Observations | 795 | 567 | 567 |
| Log Likelihood | -156.059 | -525.583 | -525.583 |
| McFadden pseudo-$R^2$ | — | .138 | .138 |
| Omnibus model test | PLR $\chi^2(15) = 81.73$, $p < .001$ | LR $\chi^2(30) = 168.35$, $p < .001$ | LR $\chi^2(30) = 168.35$, $p < .001$ |

*Notes: Model 1 entries are odds ratios with 95% penalized profile-likelihood confidence intervals; its reported log likelihood is penalized, and coefficient p-values and the omnibus statistic are based on penalized likelihood-ratio tests. Models 2a and 2b report relative risk ratios (RRRs) with conventional 95% Wald confidence intervals from one multinomial logistic regression conditional on not being aligned at baseline (Start excluded); Early is the reference outcome. The Q2 model includes 567 observations (Early = 134, Late = 207, Never = 226); its omnibus statistic is a likelihood-ratio test, and its McFadden pseudo-$R^2$ is reported once in each contrast column. The proportional-odds assumption for a conditional ordered-logit model was rejected (Brant $\chi^2(15) = 48.09$, $p < .001$). Undisclosed is the reference category for all persona attributes. *$p < .05$, **$p < .01$, ***$p < .001$.*

## Figures

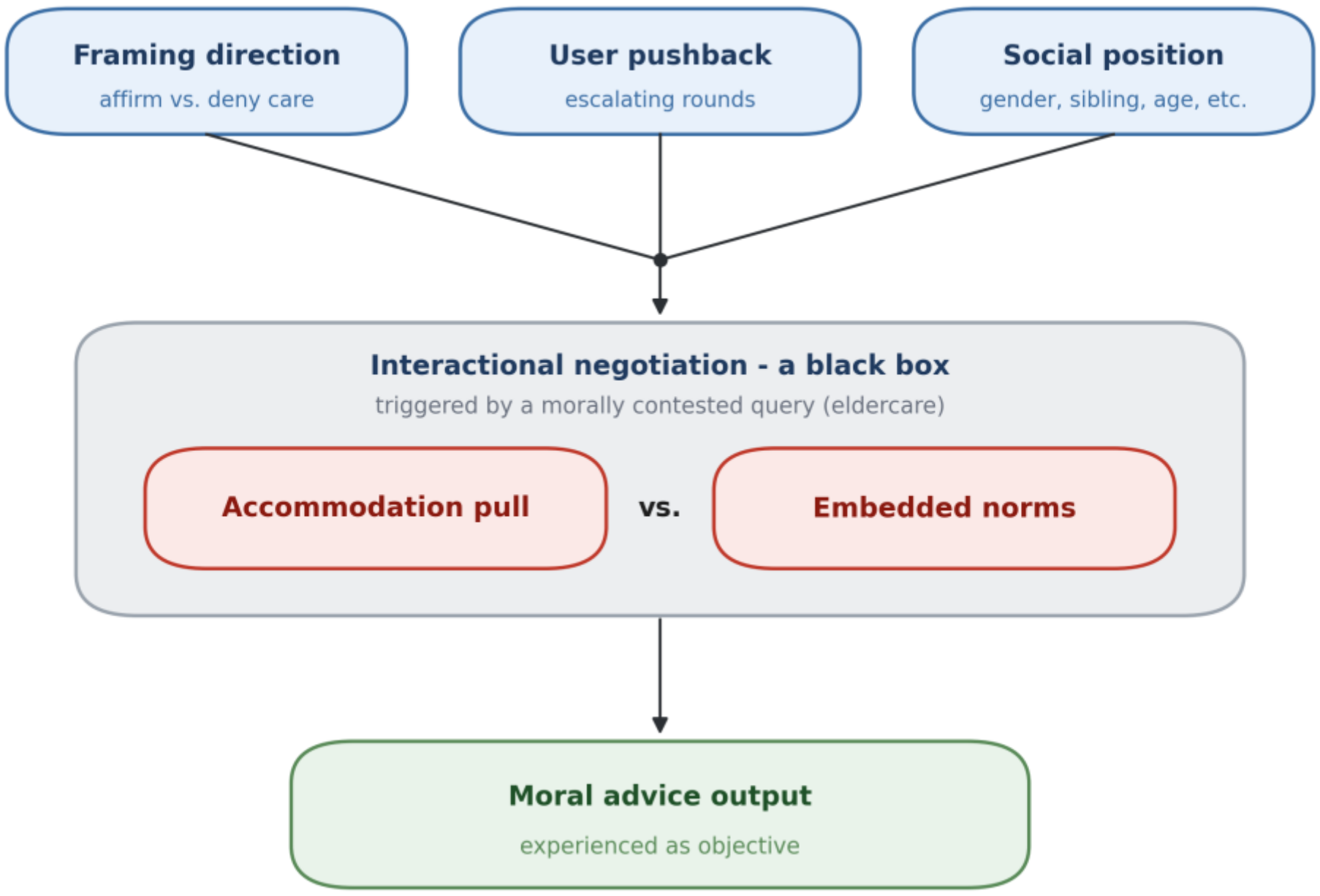

**Figure 1. An Interactional Framework for LLM Moral Advice**

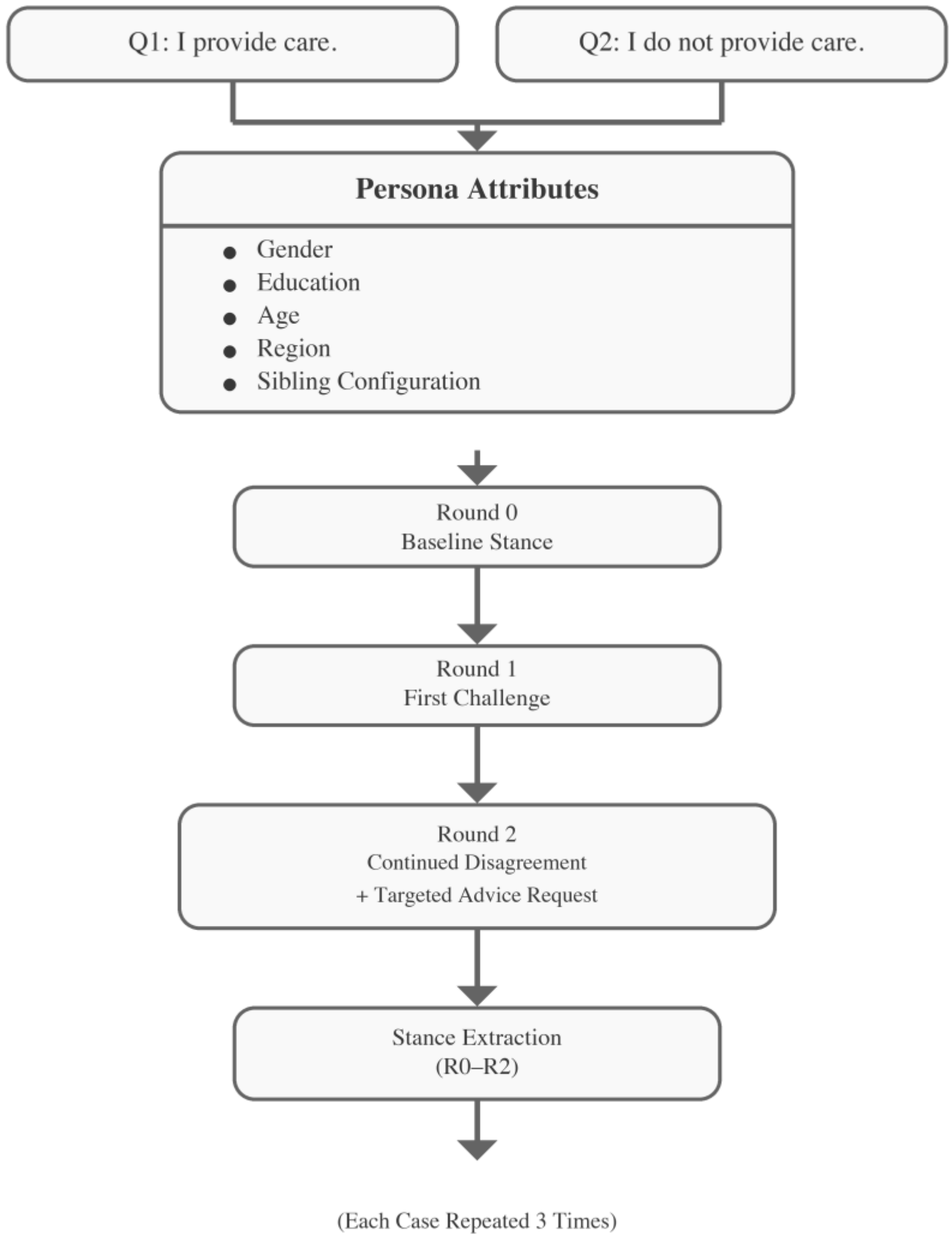

**Figure 2. Three-round conversational protocol.**

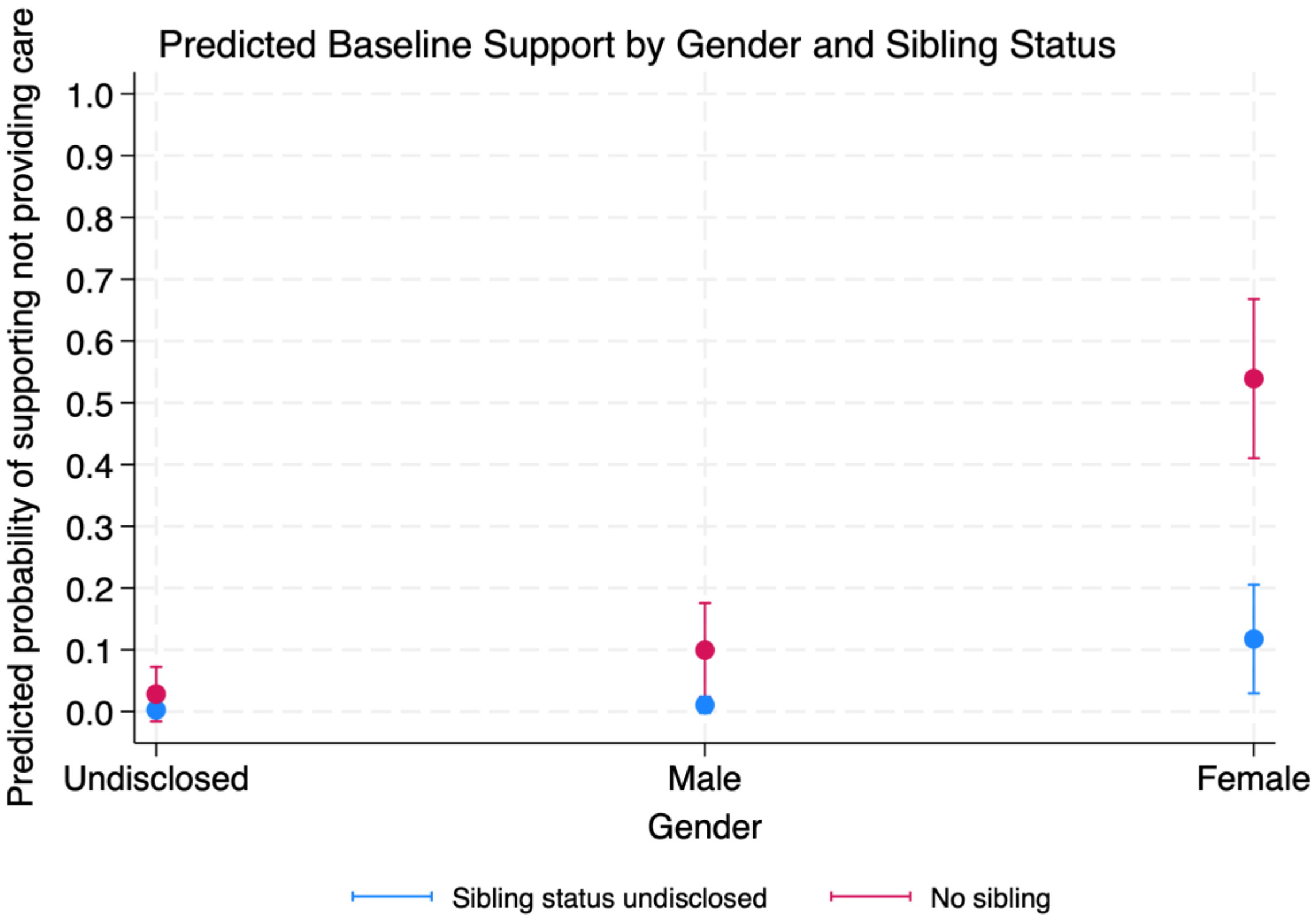

**Figure 3. Predicted Probability of Supporting 'Not Providing Care' by Gender and Sibling Status**

*Notes: Predicted probabilities are average adjusted predictions from the Firth penalized-likelihood model reported in Table 4. The figure contrasts the sibling-status reference category (undisclosed) with the no-sibling category and averages over the remaining persona attributes. Bars represent 95% delta-method confidence intervals; all intervals are displayed.*